# Optical coherent perfect absorption and amplification in a time-varying medium

Emanuele Galiffi[1]*⌄, Anthony C. Harwood[2]*, Stefano Vezzoli[2], Romain Tirole[1,2], Andrea Alù[1,3]‡ & Riccardo Sapienza[2]†

[1]*Photonics Initiative, Advanced Science Research Center, City University of New York, 85 St. Nicholas Terrace, 10031, New York, USA*

[2] *The Blackett Laboratory, Imperial College London, South Kensington Campus, SW72AZ, London, United Kingdom*

[3]*Physics Program, Graduate Center, City University of New York, 365 5th Avenue, 10016, New York, USA*

* These Authors contributed equally.

⌄ egaliffi@gc.cuny.edu

‡ aalu@gc.cuny.edu

† r.sapienza@imperial.ac.uk

**Abstract**

Time-invariant photonic structures amplify or absorb light based on their intrinsic material gain or loss. The coherent interference of multiple beams in space, e.g., in a resonator, can be exploited to tailor the wave interaction with material gain or loss, respectively maximizing lasing or coherent perfect absorption. By contrast, a time-varying system is not bound to conserve energy, even in the absence of material gain or loss, and can support amplification or absorption of a probe wave through parametric phenomena. Here, we demonstrate theoretically and experimentally how a subwavelength film of indium tin oxide, whose bulk permittivity is homogeneously and periodically modulated via optical pumping, can be dynamically tuned to act both as a non-resonant amplifier and a perfect absorber, by manipulating the relative phase of two counterpropagating probe beams. This extends the concept of coherent perfect absorption to the temporal domain. We interpret this result as selective switching between the gain and loss modes present in the momentum bandgap of a periodically modulated medium. By tailoring the relative intensity of the two probes, high-contrast modulation can be achieved with up to 80% absorption and 400% amplification. Our results demonstrate control of gain and loss in time-varying media at optical frequencies and pave the way towards coherent manipulation of light in Floquet-engineered complex photonic systems.

**Introduction**

Time-varying photonic structures have recently attracted significant interest thanks to their prospects of overcoming fundamental limitations of static photonic systems, such as reciprocity [1]-[3], absorption bandwidth limitations [4], and the fundamental constraint between loss and dispersion [5]-[6], as well as enabling nonreciprocal beamforming [7], negative refraction [8] and mirrorless lasing [9], among other effects [10]. Time-varying media, driven by an external pump, are not subjected to the constraints of intrinsic energy conservation, hence they may be used to dynamically manipulate gain and loss in otherwise Hermitian systems, i.e. to achieve wave amplification or absorption in materials with no intrinsic gain or loss.

Conventional coherent perfect absorption (CPA) [11]-[14] hinges on the coherent illumination of an absorbing scatterer by multiple beams to *spatially* maximize their overlap with material losses in the structure. At the critical coupling condition, the coupling strength is exactly balanced by the internal dissipation, enabling perfect absorption, with applications from sensors to logic gates [15]-[18] (Fig. 1a, top). The reverse process, lasing, exploits gain media in cavity resonators, whose *spatial* mode profile can be tuned to maximize the efficiency of coherent amplification by the gain medium leading to lasing by stimulated emission (Fig. 2a, bottom) [19]. However, the mode of operation of these static non-Hermitian systems is fixed by design, realizing either gain or loss depending on the properties of the medium, and is intrinsically narrowband. Indeed, because of the critical coupling condition, the smaller the intrinsic gain/loss, the higher the quality of the resonator required for either lasing or CPA.

By contrast, time-varying media can provide both gain and loss within the same structure [9],[20], with a bandwidth only limited by the speed of the modulation and without the need for intrinsic loss or gain [21]. As illustrated in Fig. 1b, the temporal version of CPA relies on the interference of the incoming forward and backward waves, with the two additional sets of waves produced by the homogeneous temporal modulation $\varepsilon(t)$. As a function of their relative phase, one can continuously switch between absorption (top case in Fig.1b) and amplification (bottom case in Fig.1b).

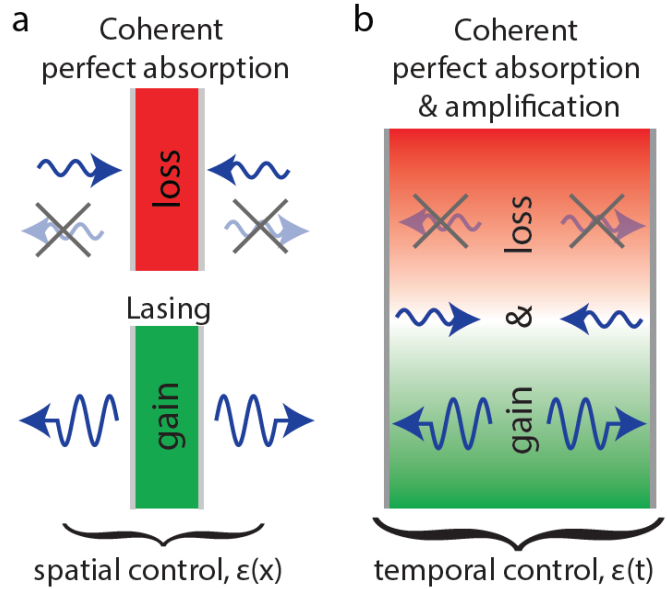

**Fig. 1:** *(a) Spatial coherent perfect absorption relies on the illumination of a lossy resonator with multiple beams in a specific amplitude and phase relation. Its time-reverse, lasing (bottom), produces coherent radiation by suitably engineering the spatial overlap of a mode with a gain medium. (b) By contrast, in a time-varying medium one can interferometrically switch between coherent absorption (crossed waves, in the red area) and gain (amplified waves in the green area), depending on the phase of the counter-propagating input beams.*

As recently demonstrated, a pulse incident on a time-varying medium can be completely absorbed when its flux is compensated by an opposite-propagating one [21]. This realizes effective collisions between optical pulses, conserving the net Minkowski momentum density, $\mathcal{P}_M = \mathbf{D} \times \mathbf{B} \propto |t|^2 - |r|^2$ (where $\mathbf{D}$ and $\mathbf{B}$ are the displacement field and magnetic induction

respectively, while $t$ and $r$ are the temporal scattering coefficients) for the respective emerging forward and backward waves [22]-[23], whereas the inelastic (i.e., absorptive), elastic (i.e., conservative), or super-elastic (i.e., amplifying) nature of the collision is dictated by the relative phase between the two incoming waves during the permittivity variation [21]. These synthetic photonic collisions were demonstrated at microwave frequencies via temporally abrupt (sub-cycle) changes in the dielectric properties of a metamaterial [21]. However, the realization of sub-cycle switching at optical frequency so far has remained elusive [24]-[27],[28].

Here we show that the coherent manipulation of gain and loss can be realized at optical frequencies by exploiting periodically modulated media, which can be modelled as driven photonic time-crystals [9]-[10],[29][30], realizing the temporal analogue of coherent perfect absorption in a thin film of Indium Tin Oxide (ITO), uniformly and periodically modulated by an optical beam. We switch interferometrically between gain and loss eigenmodes in the momentum bandgap of the modulated system by tuning the relative phase and amplitude between our probe beams, and observe absorption and amplification, achieving visibilities up to $\approx 90\%$, between $\approx 80\%$ absorption and $\approx 400\%$ amplification of a signal, under fixed input power of all beams. Moreover, our temporal CPA scheme reaches a $\approx 50\%$ modulation of the total input energy without the need for intrinsic gain or absorption, as the pump acts as an energy sink/bath. We also show that further improvements can be made by finely balancing slow and fast modulation processes in ITO. Our time-varying structure realizes highly efficient, dynamically tunable, non-resonant amplification and absorption of pulses, which may find immediate applications for telecom signal processing, pulse-shaping and photonic computing, while shedding light on the interplay between slow and fast nonlinearities in transparent conductive oxides.

**Results and Discussion**

Consider for simplicity a bulk medium undergoing a periodic square-shaped temporal modulation of its dielectric constant, with angular frequency $\Omega$ (Fig. 2a, top). The band structure of this driven photonic time-crystal can be calculated analytically [30], resulting in momentum bandgaps near the frequency $\omega_0 \approx \Omega/2$, whose width depends on the modulation depth. The gap hosts both positive and negative imaginary Floquet eigenfrequencies (Fig. 2a, bottom), respectively responsible for loss (denoted in red) or gain (green) modes [9],[29]. Dual to a spatial photonic crystal, where energy conservation results in the sole excitation of the decaying mode by a single impinging wave (in the absence of a termination), in the temporal scenario only the growing wave is excited. In fact, the conservation of Minkowski momentum implies that the inevitable production of backward waves must be compensated by an equal amount of additional forward flux, i.e., amplification, while excitation of a decaying mode would necessarily violate this conservation law. As a result of this symmetry-rooted constraint, temporal photonic crystals have been studied primarily in the regime where they operate as parametric amplifiers [29][30].

However, in a spatial photonic crystal, the growing mode simply corresponds to waves decaying from the opposite side [31]. Hence in a dual fashion, it is possible to excite the decaying mode in a driven photonic time crystal by illuminating it with counterpropagating waves (Fig. 2b, top): a rightward 'Signal' (blue wave) and a leftward auxiliary wave termed "Ancilla" (green or red wave), which we assume for now to be of identical amplitude, with relative phase $\varphi$. As the interference between the two probes produces a standing wave, all the energy density $W = (\varepsilon|E|^2 + \mu|H|^2)/2$ in the system is periodically exchanged, everywhere in space, between electric and magnetic fields, with angular frequency $\Omega = 2\omega_0$.

Simultaneously, each decrease in the modulated permittivity (Fig. 2b, middle) will increase the electric energy density, extracting energy from the modulating mechanism, while each permittivity increase will deplete it, as discussed in Refs [10],[24]-[25]. Thus (Fig. 1b, bottom), the phase between the periodic rise and fall in permittivity and the periodic storage of wave energy in the electric fields ($W_e = \varepsilon|E|^2/2$) determines the onset of gain, increasing the wave energy at each cycle (green line in Fig. 1b, bottom), or loss, depleting it (red line). Since the relative phase of the electric energy density can be set by tuning the relative phase $\varphi$ between Signal and Ancilla, this strategy enables temporal interferometric control of both gain and loss, merging CPA and lasing functionalities into a single device.

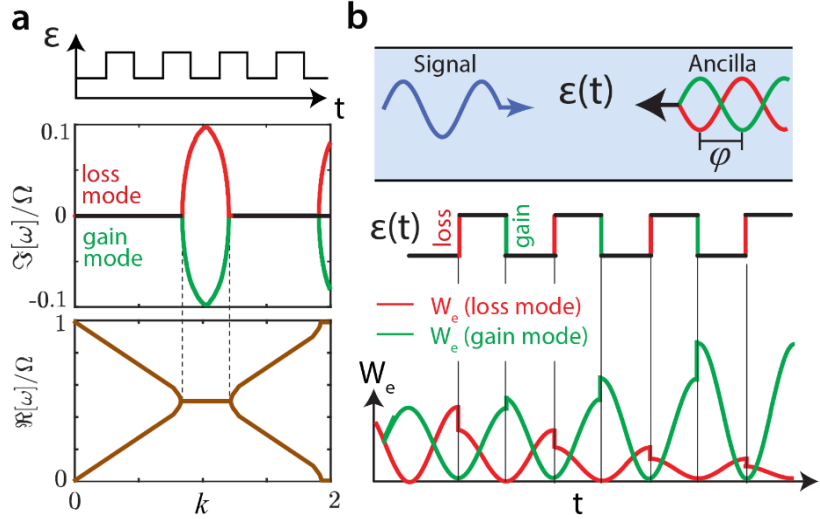

**Fig. 2:** *(a) A periodic permittivity modulation with angular frequency $\Omega$ opens a momentum gap in the dispersion of impinging light of frequency $\omega \approx \Omega/2$ into a loss mode (red, $\Im[\omega] > 0$) and a gain mode (green, $\Im[\omega] < 0$). Upon excitation by a single wave, however, only the gain mode is excited, producing equal additional forward and backward flux. (b) Conversely, by illuminating the medium with two counterpropagating beams (top), compensating forward and backward momentum flux, it is possible to select both gain and loss mode. This is because the standing wave produced by the two beams results in (bottom) an oscillation of the total energy between electric and magnetic fields. Since a permittivity increase (decrease) results in a depletion (enhancement) of the wave energy stored in the electric fields, the relative phase $\varphi$ between the two waves, and the consequent phase-locking between the permittivity modulation and the peaks and troughs of electric energy density $W_e$, determine its resonant enhancement or suppression, corresponding to the excitation of the gain or the loss mode.*

In optics, a realistic instantaneous permittivity modulation can be driven at the carrier frequency of a pump field in a material exhibiting third-order nonlinearities via the instantaneous Kerr effect [32]-[33]. Thus, upon illumination with a strong pump field of amplitude $E_{pm}$ and carrier frequency $\omega_0$, a permittivity modulation can be described as

$$\delta\varepsilon(t) \approx \beta E_{pm}^2 \approx 2\alpha \cos(2\omega_0 t), \quad (1.1)$$

which retains the same key features of the square modulation, albeit with lower efficiency (here $\beta$ is related to the $\chi^{(3)}$ of the material). As recently demonstrated [32], a wave of frequency $\omega_0$ impinging upon a periodically modulated infinite medium, without crossing spatial boundaries, may interact with the medium's momentum bandgap, no matter how small the latter may be.

Recently, unstructured thin films of transparent conductive oxides such as ITO have been shown to support efficient ultrafast modulations of their polarizability [34], leading to significant changes in Fresnel coefficients [35], efficient time-reversal [36], negative refraction [8] and time-diffraction [37]. These opportunities arise from the low carrier concentration and high melting temperatures of these semiconductors, which endow them with giant nonlinearities near their epsilon-near-zero frequency and the ability to sustain extremely large pumping intensities, enabling switching timescales of a few femtoseconds, although no agreement has yet been reached on the nature and ultimate timescale of these nonlinearities [38]-[42].

In the experiments, our amplifying-absorbing device consists of a subwavelength ITO film (310 nm, see Methods Sec. 1 and Fig. S1), over a 1 mm-thick glass coverslip, pumped at normal incidence near its ENZ frequency (1170 nm) and probed by two counter-propagating beams at the same frequency. The periodic modulation at frequency $\omega_0$ drives a periodic modulation of the dielectric permittivity at frequency $2\omega_0$, which acts as a time grating in the ITO with a narrow momentum bandgap.

The two probe beams are incident upon the device from opposite oblique angles $\theta = \pm 8°$ (Fig. 3a) and are subsequently reflected towards separate detectors. Furthermore, due to their interaction with the momentum bandgap (Fig. 3b, top), Signal and Ancilla are individually amplified in their respective forward direction, while producing a backward "phase-conjugated" (PC) wave, illustrated as smaller pink and light blue arrows in the diagram in Fig. 3a (each arrow length denotes the respective relative beam power). Importantly, the backward waves can be thought of as negative-frequency waves, excited by the permittivity modulation at frequency $2\omega_0$ via a vertical photonic transition $\omega_0 \to -\omega_0$ (Fig. 3b) [43]. The interference between PC Ancilla and reflected Signal within the momentum gap results in the coherent amplification or absorption of the Signal, measured at detector 1 (Det. 1, red). To completely erase or amplify our Signal wave under low pump intensities, we choose the intensity of the Ancilla (red arrow in Fig. 3a) to be significantly larger than that of the Signal (blue arrow).

The optical setup is shown in Methods Sec. 2 and Fig. S2. Insofar as the pump can be regarded as an infinite energy bath for the time-varying medium, which can concede or subtract energy from the two probes depending solely on the illumination scheme, and that the Ancilla is weak enough to be treated linearly, this strategy enables the efficient absorption or amplification of the Signal of arbitrary intensity purely via the appropriate tuning of the Ancilla intensity. In other words, since the pump strength only determines the amplitude of the modulation and therefore the momentum bandgap size, weak pumping only restricts the maximum bandwidth of this effect, and the minimum Ancilla intensity needed to erase the Signal. Hence, in Fig. 3 we choose a relatively low pump intensity (18 GW/cm2), which is reflected in an optimal intensity ratio between Signal and Ancilla $|E_{anc}|^2/|E_{sig}|^2 \approx 806$, ensuring that the intensity of the reflected Signal and that of the PC Ancilla at Detector 1 are comparable. Since the Signal is much weaker than the Ancilla, its amplification contribution can be neglected for such low pump intensities, such that the Signal-to-Ancilla ratio effectively quantifies the PC efficiency.

Fig. 3c shows our experimental results (red line with circles) as the ratio between the output intensity $I_1$ measured at Det. 1 and the intensity of the bare reflected Signal $I_{sig,ref}$ measured in the absence of an Ancilla, at the optimal intensity ratio between input Ancilla $I_{anc,in}$ and input Signal $I_{sig,in}$. This configuration enables interferometric switching between 400% amplification ($I_1/I_{sig,ref} \approx 4$) and $\approx 80\%$ loss ($I_1/I_{sig,ref} \approx 0.2$) of the Signal, amounting to a net fringe visibility of $\approx 90\%$. Our results are accurately matched by our linear scattering theory (blue line, see Methods Sec. 3 and SM Sec. 1 for details), which models ITO as a periodically time-modulated slab, invariant along the in-plane direction, and calculates the scattered amplitudes of the reflected Signal and PC Ancilla assuming a modulation amplitude of the background permittivity $2\alpha \approx 1.65\%$. Similar results were also obtained over a larger bandwidth, due to the non-resonant nature of the amplification process (see SM Fig. S6).

Fig. 3d shows the logarithm of the output intensity ratio $I_1/I_{sig,ref}$ as a function of both Ancilla-Signal input phase and amplitude ratio $I_{anc,in}/I_{sig,in}$, clearly identifying the relative intensity optimum shown in panel (c), and effectively defining a phase diagram for the input regimes whereby the device performs as an amplifier or as an absorber. Importantly, for Ancilla-to-Signal ratios higher than the value which produces perfect Signal absorption, this mechanism enables arbitrary amplification of the output Signal by tuning the Ancilla intensity, without varying the pump power. Through this strategy, in our experiments we reach a maximum Signal amplification of $\approx 2600\%$ by linearly varying the Ancilla power. The only limit to this amplification capability is the onset of nonlinearities that may arise once the Ancilla intensity approaches that of the pump, as in that regime the Ancilla dynamics can no longer be treated as linear. Again, measurements (Fig. 3d, left) and theoretical predictions (Fig. 3d, right) exhibit remarkable agreement for all Ancilla-to-Signal ratios considered in our experiment. Here we assume a small incoherent component of the measured intensity $I^{(net)} = \gamma I^{(inc)} + (1-\gamma)I^{(coh)}$ with $\gamma = 0.13$, coming from spurious scattering of the two probes and imperfections in the spatial and temporal overlap of the pulses at Det. 1, as discussed below.

Finally, in Fig. 3e we study the modulation visibility $v = (I_{1,max} - I_{1,min})/(I_{1,max} + I_{1,min})$ as a function of the Ancilla-to-Signal intensity $I_{anc,in}/I_{sig,in}$, by varying the intensity of the input Signal. Although we find remarkable agreement between theory (blue line) and experimental data (red line with error bars) in reproducing the optimal visibility $v \approx 90\%$ for $I_{anc,in}/I_{sig,in} \approx 806$, as well as all other Ancilla-to-Signal intensity ratios considered, the maximum visibility does not reach 100%. This can be explained by slight differences in the spatial profiles of the Signal and Ancilla beams, as well as small incoherent contributions coming from spectral difference. As typical in interferometry, the angular alignment of the two beams and their spatial overlap is critical in determining the maximum attainable modulation. Small misalignments and asymmetries in the geometry of the Signal/Ancilla are the main source of fluctuation in the measured modulation contrast (see SM Sec. 2 for more details). Moreover, our experiment is not performed in vacuum, which introduces additional fluctuations in the phase dependence of the measured output intensity at Det. 1, evident in both panel (c) and (d), which also lead to reductions in the measured visibility.

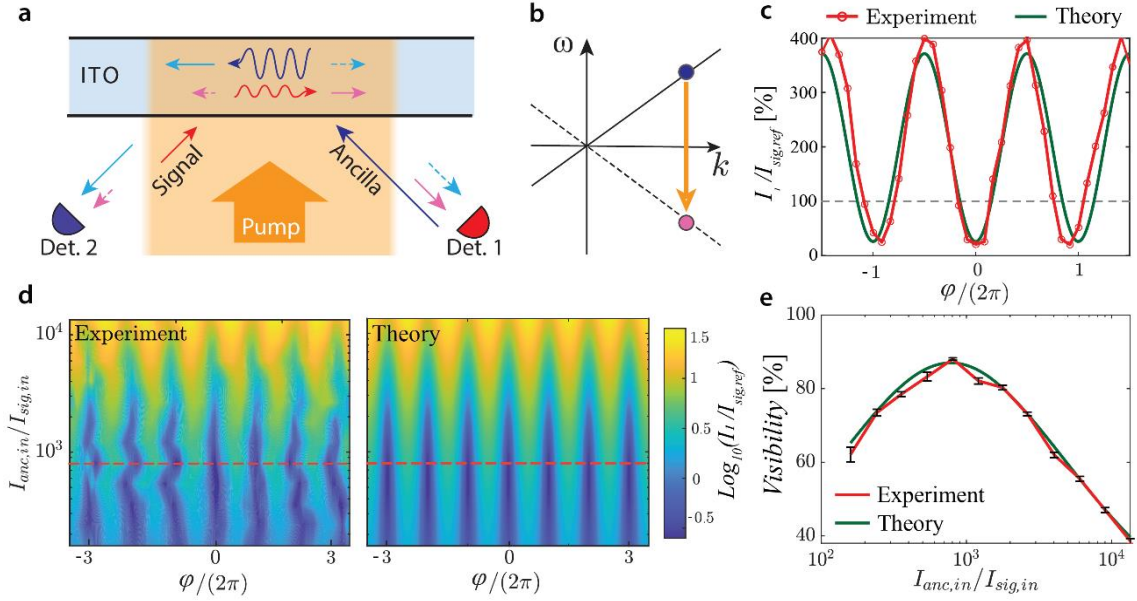

**Fig. 3:** *(a) Sketch of the setup, time modulation and illumination scheme: a subwavelength ITO slab is periodically modulated by a normally incident pump (orange). Two probe beams, a weaker beam ("Signal", red) and a stronger auxiliary beam ("Ancilla" blue) impinge on the modulated ITO from opposite angles $\theta = \pm 8°$ to the normal. Here we consider the total intensity at Det. 1. (b) The pump induces (top) a periodic temporal modulation $\delta\tilde{\varepsilon}(t) = 2\alpha \cos(2\omega_0 t)$, producing (bottom) phase-conjugated (i.e. negative frequency) waves via vertical photonic transitions ($2\alpha \approx 1.65\%$). (c) By varying the relative phase between the two probes at the optimal input Ancilla-to-Signal intensity ratio $I_{anc,in} / I_{sig,in} \approx 806$, at which the phase-conjugated Ancilla intensity matches that of the reflected Signal, we measure a net Signal amplification of $\approx 400\%$ and $\approx 80\%$ absorption (red line with circles), both accurately matched by our linear scattering model (blue line). (d) Intensity $I_1$ at Det. 1, normalized with respect to the reflected Signal intensity $I_{sig,ref}$ under pumping but without Ancilla, as a function of phase and Ancilla-to-Signal intensity ratio, showing experimentally measured (left) and theoretically predicted (right) coherent amplification and absorption, and identifying the optimal Ancilla-to-Signal ratio. (e) The interference visibility as a function of Ancilla-to-Signal ratio exhibits a maximum of $\approx 90\%$, and is accurately reproduced by our model.*

To remark how this scheme is dual to a spatial beamsplitter, in Fig. 4 we show how, in contrast with the latter (Fig. 4a), whereby a fixed amount of power is redistributed between ports, in a time-varying, spatially homogeneous structure (Fig. 4b) momentum conservation demands that both forward and backward-travelling waves undergo the same intensity modulation. To verify this prediction, Fig. 4c shows the output at both Det. 1, which measures the superposition of reflected Signal and PC Ancilla (red), and Det. 2, which measures the superposition of reflected Ancilla and PC Signal (left panel). Indeed, the power modulation at the two detectors (center panel) is nearly identical ($\approx 0.02 mW$) and follows the same dependence on the Signal-Ancilla phase delay, thus verifying this symmetry.

Expectedly, however, the visibility at Det. 2 is much smaller (only $\approx 4.5\%$) compared to Det. 1 ($\approx 61\%$), since the baseline intensity is much higher. In addition, at these pump intensities ($\approx 77 GW/cm^2$ in Fig. 4c), the average power on Det. 2 is $\approx 20$ times larger than the one measured on Det. 1. Hence the modulation of the sum of the output probe intensities (Fig. 4c, right panel), normalized to the sum of the output intensities under single-probe illumination, is less than 10%. Moreover, because of the strong unbalance between the two channels, the uncertainty in our measurements (grey band) is dominated by the power fluctuations on Det. 2 and thus is comparable with the amplitude of the modulation.

By contrast, increasing pump intensity (Fig. 4d, left) and matching the power of the input probes realizes coherent absorption and gain for *both* Signal and Ancilla (middle panel). This realizes a large modulation ($\approx 90\%$ peak-to-trough) of the *total* output power in the system (right panel), thus clearly highlighting the difference and duality of this scheme compared to spatial photonic structures. In this experiment with matched input beams, the visibility of the intensity fringes for Signal, Ancilla and their combined outputs is only $\approx 41\%$.

In principle, obtaining perfect cancellation of both waves through a photonic time grating is only possible in the extreme limit whereby the amplitude of the initial wave is negligible compared to that of the PC waves. This can be deduced from momentum conservation: destructive interference between the forward (T) and backward (R) waves produced from identical inputs clearly requires $|t|=|r|$. However, since the Minkowski momentum density $\propto |t|^2 - |r|^2$ must be conserved for each beam, $|r|^2$ can tend to $|t|^2$ only in the limit where both coefficients diverge, such that their respective amplitudes dominate over their difference.

Another competing reason for the loss of visibility in our proposed interferometric scheme is the difficulty of achieving good alignment and temporal overlap of the probe beams simultaneously on both detectors, which is considerably more delicate and challenging than optimizing for a single channel at the time without matching the input probe intensities. This can be deduced partly from the fact that the outputs at Det. 1 and 2 are different despite symmetric input and pumping conditions.

However, the most important reason for the loss of visibility is that at high pump intensities the photocarrier-mediated, non-instantaneous nonlinearities lead to macroscopic modulations of the reflected and amplified beams, which affect the spectra of Signal and Ancilla and can thus significantly impact their interference. To clarify this effect, in Fig. 5 we investigate the dependence of the gain and loss fringes on the pump power.

For strong pumping, ITO supports very large Kerr-induced index modulations that are 'slow', i.e. roughly following the pump envelope and subject to slow relaxation times of hundreds of femtoseconds [35][37][40], as opposed to the effectively instantaneous ('fast') modulation discussed so far [44]. These slow modulations can significantly shift the plasma frequency of the material, and thus its refractive index, significantly changing the reflected probe power, as well as red-shifting and broadening its spectrum [35][37][40]. The most immediate consequence is a significant reduction in steady-state reflectance, which limits the extent to which the probes can be efficiently amplified. This reflectance variation is modelled in our theory by including a redshift in the steady-state plasma frequency $\delta_{\omega_p} = \beta_{\omega_p,slow} I_{pump}$ [28][35][37][40], with a proportionality constant $\beta_{slow} = -0.0003 cm^2/GW$. As evident from Fig. 5a (red, right axis) this model (solid line) captures well the observed reflectance (dashed line with circles) dependence on the pump intensity, up to $I_{pump} \approx 150 GW/cm^2$, while higher-order effects may be needed to model even higher pump intensities.

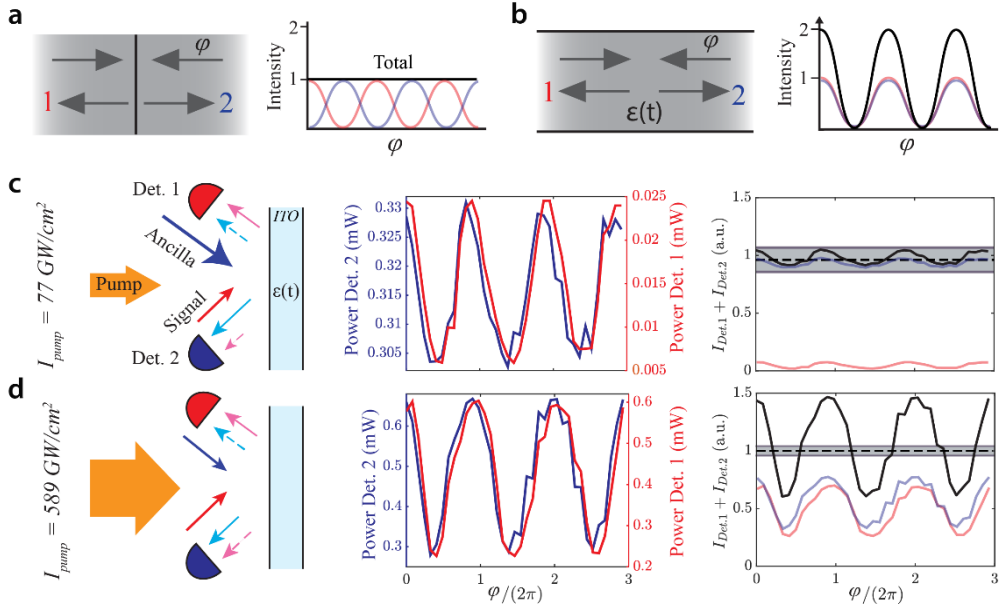

**Fig. 4:** *(a) A conventional beamsplitter redistributes a fixed amount of power from multiple impinging waves between its output ports. (b) By contrast, a time-varying medium can combine incoming waves such that all of them are suppressed or enhanced by the same amount, obeying momentum conservation, rather than energy conservation. Thus, by using (c) Ancilla and Signal of very different power (left diagram), we can always find a Signal-to-Ancilla ratio that cancels or amplifies the Signal (red y-axis in center panel) for a given pump power, while causing the exact same intensity modulation of the Ancilla (blue axis), as expected by in-plane momentum conservation. However, note how the relative modulation is much smaller for the reflected Ancilla (Det. 2) than for the reflected Signal (Det. 1), as the power at the former is ~ 20 times higher. This results in (right panel) a weak modulation of the relative total output power, obtained by summing the intensities in the two output channels (shaded lines), comparable to the measured uncertainty (shaded area). (d) Increasing the pump power (left) allows us to work with similar input Signal and Ancilla intensities, resulting in (center panel) identical output amplitudes and baselines for both outgoing Signal and Ancilla, and a strong modulation of the total output probe intensity (right).*

The PC efficiency (left axis, blue) is also reproduced accurately by our model (solid line), matching the experimental measurements up to $I_{pump} \approx 60 GW/cm^2$ (dashed lines with circles) agrees well with the theoretical prediction, which assumes a modulation amplitude $\alpha = \beta_{inst} I_{pump}$, up to pump intensities $\approx 60 GW/cm^2$, where, for this instantaneous process we find $\beta_{inst} = 0.002 cm^2/GW$. Even better agreement for high pumping may be obtained by using models that can combine both fast and slow timescales [45], although we postpone such deeper theoretical investigation to a dedicated computational study. Interestingly, for very high pump powers the amplitudes of reflected and PC waves appear to cross, implying that it may be possible to leverage the interplay between fast and slow modulations to realize a perfect absorber for identical probes.

More subtly, the impact of the stronger pumping also affects the visibility of the gain-loss fringes. As shown in Fig. 5b, the visibility is significantly reduced for intensities larger than $\approx 100 GW/cm^2$. Surprisingly, however, it increases again for $I_{pump} > 300 GW/cm^2$. To understand this, in Fig. 5c we show spectrograms of the (top) PC Ancilla and (bottom) reflected Signal as a function of pump power. We observe that the onset of the redshift and spectral broadening induced by the slow ITO modulation on the reflected Signal occurs at lower pump intensities than for the PC Ancilla, similarly to what observed recently in another ITO-based time-varying mirror[37][40]. This frequency shift clearly results in a loss of amplitude and phase coherence between the two beams, thus reducing significantly the visibility (see SM Sec. 3 for details).

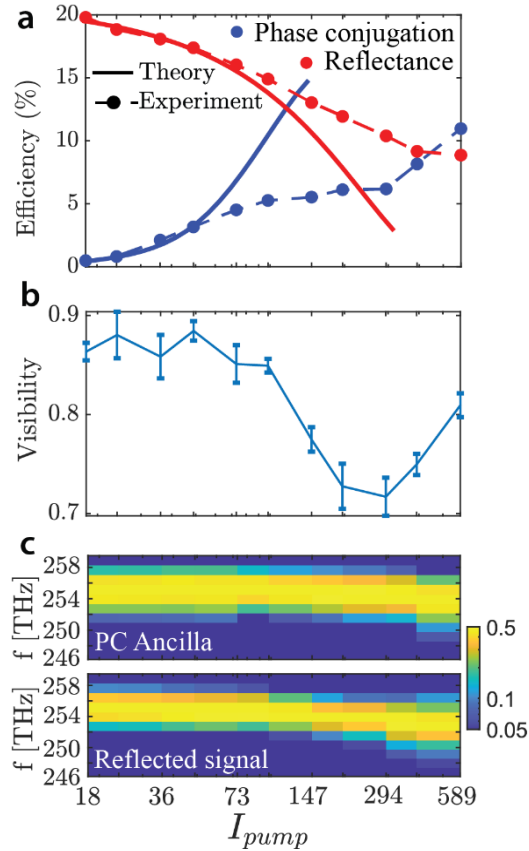

*Fig. 5: (a) The measured (dashed lines with circles) and theoretically predicted (solid lines) efficiency of phase conjugation (left axis, blue) and the modulated reflectance induced by pumping (right axis, red lines), plotted as a function of pump intensity (log-scale), match accurately for weaker pumping, while the agreement worsens for pump powers above $\approx 60 GW/cm^2$ for the phase conjugation, and above $\approx 150 GW/cm^2$ for the reflection modulation. (b) The fringe visibility drops significantly for pump intensities $\approx 100 GW/cm^2$, increasing again for intensities above $300 GW/cm^2$. This is explained by (c) the redshifts undergone by (top spectrogram) the reflected Signal and (bottom spectrogram) phase-conjugated Ancilla: the former is significantly redshifted for lower pump powers than the latter, thus spoiling the coherence needed to achieve strong visibility fringes. As the phase-conjugated Ancilla is also redshifted for even higher pump intensities, some coherence is restored.*

Indeed, this coherence is partially restored for even higher pump powers, as the spectrum of the PC Ancilla undergoes a similar redshift to that of the reflected Signal. Interestingly, this asymmetry suggests that the interplay between fast and slow nonlinearities may circumvent the symmetry constraint between parametric amplification and PC, potentially enabling nonreciprocal phenomena dual to those shown by combining nonlinearity and spatial asymmetry in Ref. [46]. Finally, our results at high pump power clearly show that our photonic time grating analogy is strictly limited to the perturbative regime of small permittivity modulations. However, thanks to the current research efforts towards the realization of an efficient photonic time crystal [28], supporting large and arbitrarily fast modulations, it is possible to envision the improvement of our current scheme to realize broadband coherent perfect absorption and amplification.

*Conclusions*

To conclude, we have demonstrated theoretically and experimentally how periodically modulated structures provide an alternative route to optical coherent perfect absorption and amplification, achieved via interferometric switching between the excitation of the gain and loss modes of an effective driven time crystal formed by an optically pumped, subwavelength film of Indium Tin Oxide. We achieve this by coherently illuminating the structure with two opposite probes: a stronger auxiliary probe pulse (Ancilla) interacts with the modulated structure to produce phase-conjugated waves, which strongly suppress or amplify a weaker counterpropagating pulse (Signal). Crucially, our strategy uses two probe pulses of different intensity, which only experience the parametric response of the time-varying medium, allowing this effect to be possible with an arbitrarily weak pump. Finally, by tuning the pump strength, we can manipulate the optimal ratio between Signal and Ancilla required to produce optimal coherent perfect absorption.

Our results open a new vista on the opportunities for optical coherent wave control phenomena in time-varying media, clarifying the importance of probe illumination protocols to achieve gain or loss in photonic time-crystals, and showcasing how these systems can effectively induce non-conservative interactions between optical pulses despite their linearity in the probe dynamics. These results may be extended in the future to incorporate parametric phenomena in more complex structures, such as complex cavities and photonic crystals, and may be extended to nonreciprocal scenarios. In addition, exploring the regime of extremely strong pumping may shine new light on the interplay between fast and slow linearities in transparent conducting oxide, unveiling new degrees of freedom for wave control with time-varying media.

**Methods**

*1. Sample characterization*

The sample consists of 310 nm of commercially available Indium Tin Oxide (from Präzisions Glas & Optik GmbH) deposited upon 1.1mm of glass, nominally identical to that used in Ref. [34] of the main text. In Fig. S1 we characterize the sample by linear ellipsometry, which gives a plasma wavelength $\lambda_p = 597$ nm, a background permittivity $\varepsilon_\infty = 4.08$ (i.e. an epsilon-near-zero wavelength $\lambda_{ENZ} = \sqrt{\varepsilon_\infty}\lambda_p = 1196$ nm). The electron scattering rate is $\gamma = 0.13 fs^{-1}$. Under the pumping conditions used in Fig. 3 (pump power $\approx 18 GW/cm^2$), we find a slight relative redshift of $\approx 0.9\%$ in the plasma frequency, which is well-known to occur from previous literature (see e.g. Refs [35][37][40]), and is used in the theory to reproduce the data in Fig. 3.

   *2. Experimental setup*

The optical system (Fig. S2) is a dual-probe variant of a degenerate pump-probe experiment. The pulses were generated by a PHAROS (Light Conversion) solid-state laser coupled with an ORPHEUS (Light Conversion) optical parametric amplifier. The subsequent pulses feature a nominal FWHM of 225 fs and could be tuned throughout the range 1070 nm to 1370 nm. The OPA output beam is first divided by an 80:20 beam splitter to generate the high-intensity Pump beam (yellow line) and the probe light, which is later split equally and further attenuated to create the low power Signal (red line) and Ancilla (yellow) beams. In contrast with conventional CPA experiments [12], where counter-propagating beams are used, all three beams are incident upon the same face of the sample; their transmission through the sample is neglected in this experiment, and we focused on reflected and phase-conjugated beams instead.

As depicted in Fig. S2, the Pump beam arrives at normal incidence and the Ancilla and Signal beams arrive symmetrically from either side of the pump, at an angle of 8 degrees to the normal. Using distinct lenses for the probe and pump beams we control the gaussian spatial profile of the beam spots at the sample, which can also be verified via an imaging camera. The pump beam is focused to a spot size of approximately 80 μm, whereas the probe beams were focused to 40 μm, ensuring a spatially nearly uniform time modulation of both probe beams. With the beams spatially superposed, these pulses were synchronized and temporally tuned with the variable delay stages included in the beam paths of the Ancilla and Pump. Additionally, the incident powers of all three beams were tuned using variable neutral density filters.

The interaction of Signal and Ancilla with the periodic modulation of the permittivity induced by the Pump generates backward propagating beams that are separated from the co-axial incident probe beams with 50:50 beam splitters and interfere at detectors 1 and 2. Spectrometers (OceanView FlameIR or Princeton Instruments NIRvana) were used to analyze the output Signal, while scanning the relative phase on Ancilla and Signal by moving the delay stage between the two probes with a step of 50 nm.

   *3. Theoretical model*

The key assumption of the model is that the plasma frequency or the background permittivity are assumed to be modulated sinusoidally at a frequency chosen as twice the modulation frequency of the pump (and, in our degenerate experiment, of the probe). The method is based on transfer matrices, but it allows for multiple frequencies, coupled to each other via the time-modulation, which enters Maxwell's equations through both terms in the displacement field (background permittivity and plasma frequency), although in this work we only modulate periodically the background permittivity.

An illustration of the scattering problem is shown in Fig. S3. Incoming ports encompass amplitudes $E_{1,-}^{(n)}$ and $E_{3,+}^{(n)}$, which are all set to zero with the exception of the fundamental harmonic in the "1" subspace $E_{1,-}^{(0)}$, which propagates in the negative z-direction towards the slab. Boundary conditions for the $E$ and $H$ field are applied frequency-wise at the two boundaries of the ITO slab of thickness "d". The model assumes a modulation amplitude and calculates the amplitudes of all outgoing harmonics $E_{1,+}^{(n)}$ and $E_{3,-}^{(n)}$ into the respective subspaces, including phase-conjugated waves $E_{1,+}^{(-1)}$ and reflected waves $E_{1,+}^{(0)}$. The amplitude of the periodic permittivity modulation is the key fitting parameter. All details and derivations are included in the Supplementary Material. As the system is a linear time-varying one, scattered waves produced by Signal and Ancilla are calculated independently and combined a posteriori to form the complete outgoing waves.

**Acknowledgments**

E.G. acknowledges funding from the Simons Foundation through a Junior Fellowship of the Simons Society of Fellows (855344/EG). A.C.H acknowledges support from the Val O'Donoghue Scholarship in Natural Sciences. A.A., R.T., E.G. were partially supported by the Office of Naval Research and the Simons Collaboration on Extreme Wave Phenomena. R.S., S.V. acknowledge support from UKRI (EP/Y015673).